# A systematic review of geospatial location embedding approaches in large language models: A path to spatial AI systems

*Sean Tucker*

Geospatial Location Embedding (GLE) helps a Large Language Model (LLM) assimilate and analyze spatial data. GLE emergence in Geospatial Artificial Intelligence (GeoAI) is precipitated by the need for deeper geospatial awareness in our complex contemporary spaces and the success of LLMs in extracting deep meaning in Generative AI. We searched Google Scholar, Science Direct, and arXiv for papers on geospatial location embedding and LLM and reviewed articles focused on gaining deeper spatial "knowing" through LLMs. We screened 304 titles, 30 abstracts, and 18 full-text papers that reveal four GLE themes - Entity Location Embedding (ELE), Document Location Embedding (DLE), Sequence Location Embedding (SLE), and Token Location Embedding (TLE). Synthesis is tabular and narrative, including a dialogic conversation between "Space" and "LLM." Though GLEs aid spatial understanding by superimposing spatial data, they emphasize the need to advance in the intricacies of spatial modalities and generalized reasoning. GLEs signal the need for a *Spatial Foundation/Language Model* (SLM) that embeds spatial knowing within the model architecture. The SLM framework advances Spatial Artificial Intelligence Systems (SPAIS), establishing a Spatial Vector Space (SVS) that maps to physical space. The resulting spatially imbued Language Model is unique. It simultaneously represents actual space and an AI-capable space, paving the way for AI native geo storage, analysis, and multi-modality as the basis for Spatial Artificial Intelligence Systems (SPAIS).

Keywords**:** geospatial location embedding; large language model; spatial language model; space as a language; spatial artificial intelligence system

**Key policy highlights**

1. Standardize Spatial Vector Space (SVS): Develop international standards for terrestrial and extraterrestrial SVS, for consistent and interoperable geospatial AI applications. This facilitates global collaboration in spatially related initiatives.
2. GeoAI Research Prioritization: Allocate resources to explore GeoAI native modalities, advancing AI-to-geospatial features awareness. This will enhance AI's role in navigation, exploration, and automation of terrestrial and extraterrestrial systems.
3. Ethical GeoAI Frameworks: Implement ethical frameworks governing the collection and use of spatial data, addressing privacy, security, and responsible AI usage. to maintain public trust as GeoAI systems as they become more socially integrated..

## Introduction

Space has always been complex to represent. Whether it is a sphere on flat paper or arcs as lines - we have always contorted space to discrete estimates of the trueness of our complexities. We strive for greater accuracy in representations to increase the validity of our urban planning, environmental, socio-economic, and other spatially related insights. Our ever-increasing activities and references to our lives are compressing more things into our space, making our spaces larger and more complex (Batty and Xie 1994; Leszczynski and Crampton 2016). The geospatial world has escalated its call for a partnership with Artificial Intelligence (AI) to improve the meaningful storage, assessment, and evaluation of Space (Smith 1984; Openshaw and Openshaw 1998; Janowicz *et al.* 2019; Liu and Biljecki 2022; Das 2023). The synergy of Geospatial Intelligence and Artificial Intelligence is - GeoAI. In the last year, 2022 – 2023, much GeoAI focus is on using the tried and proven 'meaning extraction' capabilities of Large Language Models (LLM) to extract deeper geospatial awareness (Mai *et al*. 2022; Bhandari *et al*. 2023; Fernandez and Dube 2023; Hu *et al.* 2023; Mai *et al*. 2023; Manvi *et al*. 2023).

Geospatial-LLM mergers are simultaneously successful and frustrating (Bhandari *et al*. 2023). Attempts to embed geospatial meaning into LLMs reveal a discord, or unnatural language, between LLMs and spatial meaning (Liu and Biljecki 2022). There are *vocabulary* Gaps between LLMs and foundational geospatial aspects, such as – location, modalities, and calculations (Yan *et al.* 2023; Li *et al.* 2023; Salmas *et al.* 2023). The result is difficulty in translating consistent geospatial meaning using LLMs. We could say that - 'space is lost in LLMs, and LLMs are lost in space.' However, the GLE investigations have yielded insights into developing a natural Space-LLM dialog (Balsebre *et al.* 2023; Ling *et al.* 2023; Fernandez and Dube 2023). The primary revelation is that we must align language models' mode, modality, and math to *speak better Space* (Li *et al.* 2023).

Historically, spatial alignments are typical for burgeoning geospatial eras. As Herring (1991) and Goodchild (2009) indicated and have proven true through many geospatial eras and sub-eras such as - ubiquitous GPS use, GIS emergence, and Geo-design – *the success of a new geospatial paradigm results from a clear conversation between the technology and the domain, where the technology finds a natural way to speak the language of space*, or we could say – speak space as a Language (SaaL). A touchstone example is - the development of the mathematical language of social and economic gravity to gain insight into the cellular level of urban geography (Batty and Xie 1994; Batty *et al*. 1999). In other words, the Space-LLM approach proves natural and implementable when built from a spatial viewpoint (Goodchild and Janelle 2004; Kuhn 2012; Xing and Sieber 2023). Space and LLM need to have a spatial conversation.

If we imagine a Herring (1991), Batty *et al*. (1999), and Goodchild (2009) (HBG) contrived conversation between Space and LLM, it would be as follows:

**Space:** 'LLM, how do you *demonstrate such deep knowing*?'
**LLM:** 'My math – my Arithmetic Embedding and Vector Space.'
**Space:** 'Arithmetic Embedding? Vector Space? How so?'
**LLM:** 1. 'Arithmetic Embedding uses Vectors to tell closeness, direction, and intention.'
2. 'Vector Space is where I put all these items related to each other.'
**Space:** 1. 'I also need to know closeness/direction, and intention!
Can my vectors be your vectors?'
2. 'My Space also puts items in their place.
Can my vector space be your vector space?'
(… to be *continued*…).

The literature reveals that this is the current stage of the Space-LLM conversation. Space wants to use LLM to speak, and LLM wants to speak space, but there are questions about calculating the meaning of items, boundaries, and relationships, which calls for a model for speaking space – an SLM. GLEs attempt to answer such questions by embedding geospatial location and descriptive data into an LLM, which helps LLMs assimilate and explain geospatial information.

This GLE review shows we are at the crossroads of GeoAI *representation and expression approaches* – how AI can represent and express space. We will see how the SaaL conversation begs us to extend representation and expression awareness to a deeper level of embedding. We posit that a review of GLE approaches also reveals an approach for AI to speak SaaL. We avoid a performance-centric viewpoint because we believe performance-centrism leads to underperformance of the tried and proven distillation of HBG – Perceive space correctly, and practical performance will come. An SLM will extrapolate on and deepen current spatial insights.

This paper examines how we use LLM Location Embeddings to try to speak 'Space.' Can we listen more closely to space? We will also extract lessons that make SaaL possible. Our three questions will traverse the Space-LLM conversation.

**Q.1**. What improvements in geospatial knowing come from using GLEs?

**Q.2.** What Gaps in Geospatial-LLM do GLEs reveal?

**Q.3.** How could we model *SLM for native GeoAI*?

Our *method* involved - a targeted search and synthesis that led to GLE insights.

## Method

### *Search strategy*

We searched Google Scholar, Science Direct, and arXiv for articles on how LLMs extract geospatial 'knowing.' The phrase chosen was 'geospatial location embedding LLM.' The date of the search is November 21, 2023. There were 304 results. Justification for the search is as follows - *Geospatial* is standard terminology in geography and geoscience literature; *Location Embedding* is the term the geospatial corpus uses to indicate Large Language Model embedding; LLM is the standard abbreviation used in Large Language Model papers and helps weed out papers that use all three words throughout but do not refer to AI LLMs. This search synthesizes literature focused on using fine-tuned LLMs for geospatial insights. A large corpus of recent papers reveals a heightened and accelerated awareness of the gaps and potential for native GeoAI.

We did a three-stage review that progressively filtered toward the relevant literature – 1. A title and summary review from the article title and summary; 2. An Abstract review that was a thorough read of the article abstract; and 3. A paper review was a comprehensive review of the paper to assess its alignment with the selection criteria, placing great emphasis on the *'In this paper'* and *'goals'* section. During the Abstract review, 'types' of location embedding emerged as a clear theme per domain effectiveness. To ensure furthering the Space-LLM conversation, we focused on papers addressing implementation. The selection criteria were – 'practical geospatial location embedding with LLM.' *Practical* means practice, theory, or modeling of an approach. *Geospatial location embedding with LLM* means the paper showed how geospatial 'knowing' could be impacted by embedding spatial information into an LLM. We excluded articles that did not meet the full selection criteria. We also excluded papers that focused solely on spatial graph data techniques. We wanted to focus on using LLMs to extract spatial meaning used in Natural Language Processing (NLP) and Generative AI for meaning extraction. We also wanted to determine if the GLE efforts for aligning Geo-space with LLMs are gaining meaningful momentum or reaching their terminal velocity.

Thirty papers passed the title and summary review. Eighteen papers passed the abstract review, and sixteen papers passed the full paper review. The fact that many of these papers cite each other reveals the building of a specialized, interconnected cohort of experts synergizing on the Geospatial-LLM trajectory. Figure 1 shows the – *GLE Distillation - From Search to Selection*.

Figure 1. GLE distillation – From search to selection

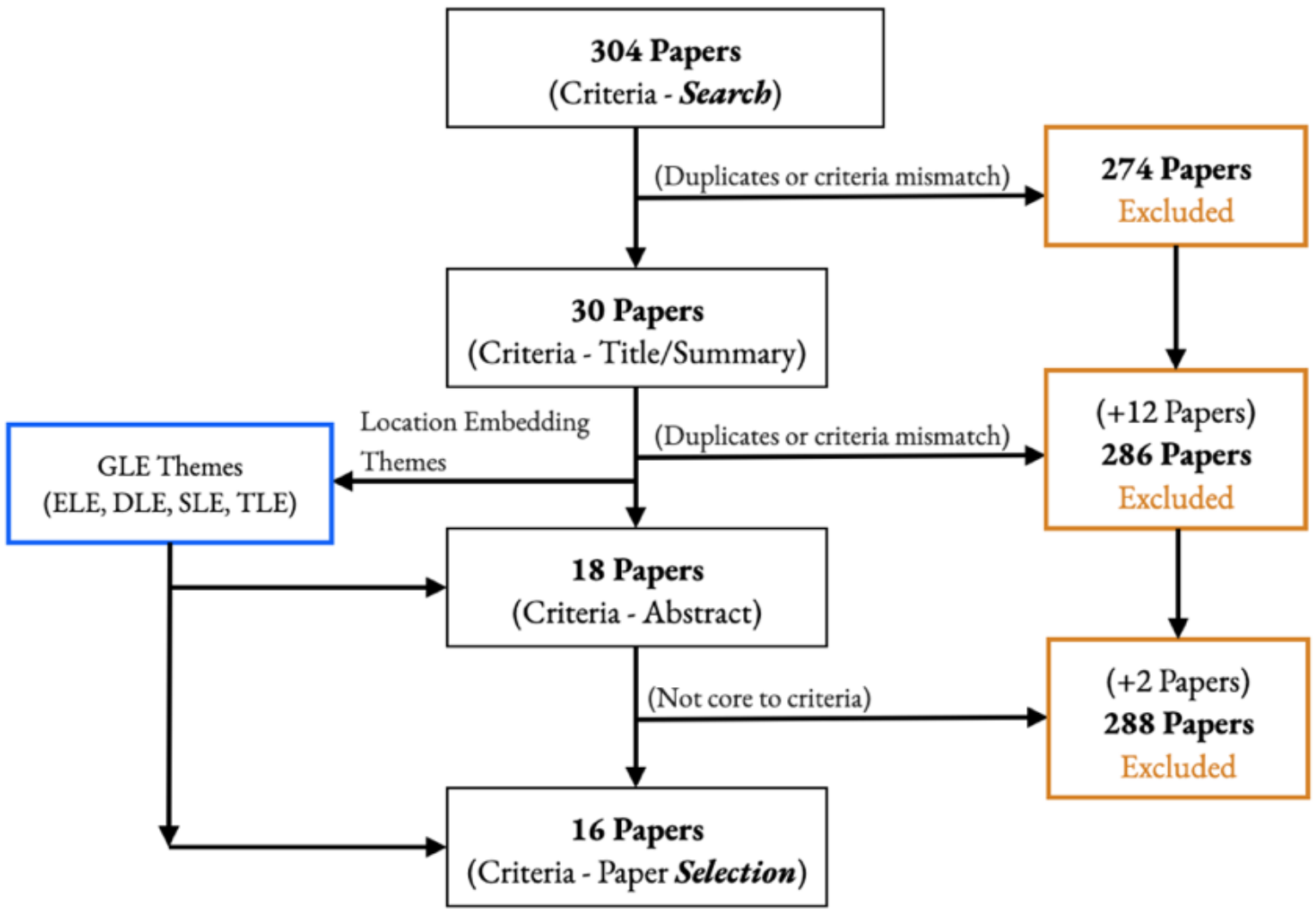

GLE distillation was done in 3 phases, yielding 16 papers on 4 GLE themes.

***Data synthesis – Reveals Geospatial Location Embedding (GLE)***

The Title and Summary Review utilized the title and summary details. The Abstract Review followed, revealing four major GLE themes – Entity Location Embedding (ELE) – embedding of *entities* such as a point of interest; Document Location Embedding (DLE) – embedding a document with spatial information about a place or an event at a place; Sequence Location Embedding (SLE) – embedding continuous features such as roads in a city, and Token Location Embedding (TLE) – embedding geospatial data such as coordinates within in the LLM. The abstract review was compared and contrasted against Generative AI prompt engineering, with ChapGPT 3.5 and Claude, using the prompt – "Evaluate the abstracts against the following approaches ... ELE includes embedding geospatial items, objects, and places as entities - for example, 'The Smithsonian' as a building or a location … DLE includes … SLE includes … TLE includes." Differences led to revisiting the abstracts to note three more documents that referenced SLE and TLE. Then we performed a full Paper Review via a thorough reading and coding using MAXQDA to verify embedding types and criteria match. Appendix *Table A.2* shows the theme selection criteria, final articles, and their associated embeddings.

***GLE approaches***

***ELE*** includes embedding geospatial items, objects, and places as entities - for example, 'The Smithsonian' as a building or a toponym. ***DLE*** includes documents and external data embedded in the LLM through augmentation - such as a brochure on The Smithsonian, containing pictures, text, location-related data treated as augmented data, or related data from Open Street Maps (OSM). ***SLE*** includes continuous geospatial feature data that form sequences or networks, such as roads around The Smithsonian from an OSM source. ELE and SLE can be combined with DLE to utilize Retrieval Augmented Generation (RAG), where RAG uses documents or other data to augment the LLM's data and assessment. ***TLE*** includes embedding spatially explicit data such as longitude, latitude, altitude, also, relative distance data, or text referring to spatial locations or items. TLE helps an LLM to see spatial values as tokens with fundamental language meaning. *Table 1* shows entity and sequence embeddings are always involved, document embeddings are often involved, and token embeddings are rare and complex. The paper titles also reveal a stronger connection between modality-specific papers and a focus on sequence embeddings, such as 'CityFM …' by Balesbre *et al.* (2023) and 'Urban profiling …' by Yan *et al*. (2023). GLEs do not address the high-dimension embedded vector space –of an LLM that captures meaning. Geospatially speaking, this would be a Spatial Vector Space (SVS), akin to a Spatial Reference System (SRS).

A *discussion* of the details reveals – GLE-LLM improvements, gaps, and trajectory.

## Discussion

*ELE*s focus on locations tied to entities in various applications, such as semantic queries, urban planning, and geocoding (Balsebre *et al*. 2023; Fernandez and Dube 2023; Ji and Gao 2023; Li *et al.* 2023; Yin *et al.* 2023). Entities, serve as a foundational building block for geospatial understanding as a single reference to many modal types. While ELEs fall short of the precision seen in conventional GIS methods, they highlight the potential of vector-based downstream modality.

*SLEs* are employed for distance, routing, and multi-modal applications in fields like imagery analysis and urban planning (Das 2023; Jin *et al*. 2023; Luo *et al*. 2023; Salmas *et al*. 2023; Yan *et al*. 2023). While SLEs provide a high-level approximation of spatial meaning, their effectiveness depends heavily on domain specialization of storage and training, making them adept at specialized domain representation but weak at generalization (Ling *et al*. 2023).

DLEs in 9 of the 16 papers leverage external data like OpenStreetMaps to enhance LLM language processing. Aimed at improving location attributing dynamically, DLEs use techniques such as Retrieval Augmentation (RAG) for better spatial generalization (Manvi *et al.* 2023; Unlu 2023). RAG's referencing to relevant data enhances spatial awareness and understanding. While this infuses LLMs with spatial context, it does not imbue them with fundamental spatial awareness.

TLEs, in 5 of the 16 papers, involves using the LLM to do Token Location Embedding LTLE to reference spatial semantic meaning. This provides an interpreted rather than intrinsic grasp of location data (Ji and Gao 2023; Fernandez and Dube 2023). Das (2023) calls for deeper imbedding for intrinsic understanding of classifications and relationships, which we posit to be the embedding that creates a language model - Vector Token Location Embedding (VTLE).

### *GLE-LLM effectiveness (Q.1.)*

ELEs effectively capture semantic relationships between entities and their locations, contributing to deep toponyms, entity reference and geoparsing capabilities (Ji and Gao 2023; Xing and Sieber 2023; Li *et al*. 2023). ELEs aim to make location intrinsic to LLMs, which we call Location Insight (LI).

SLE's effectiveness is achieved through domain-specific approaches, enhancing an LLMs' ability to identify shapes and spatial relationships (Ji and Gao 2023; Jin *et al*. 2023; Das 2023). SLEs aim to make structure intrinsic to LLMs, which we call Modality Insight (MI).

DLE's effectiveness stems from extracting meaningful text corpora, incorporating techniques like RAG to handle geographic questioning. DLEs use Referential Insight (RI) to improve effectiveness (Ling *et al*. 2023; Mai *et al*. 2023; Manvi *et al.* 2023; Yan *et al*. 2023).

TLEs, including LLM Token Location Embeddings (LTLE), increase location and representation awareness, addressing contextual issues (Fernandez and Dube 2023; Ji and Gao 2023). LTLEs provide Context Insight (CI) by defining dimensions and determining context and meaning.

### *Common success and gaps*

The common success in GLE-LLM lies in intrinsic modeling, aligning with the nature of spatial location, modality, references, and context. The efficiency of these models aligns with domain-specific approaches but highlights challenges with generalization across domains (Balsebre *et al*. 2023; Xing and Sieber 2023; Ling *et al*. 2023; Mai *et al*. 2023; Yan *et al*. 2023). Aligning mode, modality, and math with the intended knowledge domain enhances the LLM's spatial language capabilities. The commonalities in efficiency align with shared gaps, emphasizing the importance of tailored approaches in GLE-LLM integration.

## GLE gaps (Q.2.)

*The common challenges* of GLE LLMs are

(1) *Mode* – What to use to store spatial items (Fernandez and Dube 2023; Mai *et al*. 2023)
(2) *Modality* – How to represent different features and shapes (Balsebre *et al*. 2023)
(3) *Math* – How to calculate locations and relationships (Han *et al*. 2023; Ji and Gao 2023)

Mode refers to the type of values we use to store features and the vector space of a model – for example, should we store items using numbers for latitude, longitude, and altitude or words that describe things? Modalities are the data and features represented by modes. Math refers to arithmetic formulas applied to the modes and modalities to represent and explain space.

Mode will facilitate or limit how we express essential spatial elements. For example, the phrase 'close to Washington, DC' is not as precise nor as mathematically available as the coordinate location Lat 38.890435, long -77.030052, alt 137, for the Smithsonian. However {38.890435, -77.030052, 137} may benefit standard vector database storage more. Then, modality impacts feature representation, for example - The Smithsonian's physical properties, including its shape, size, and borders, using multiple entries of the number representations. Then, mathematically, the last number representation is more mathematically convenient since it only uses numbers, and the relative positions of the numbers indicate their representation and relationship.

The mode-modality-math challenges can be overcome by helping Space and LLM mode, modality, and math to map clearly to each other.

A review of the Space-LLM conversation reveals that the GLEs attempt to address the first premise of the Space-LLM conversation – Arithmetic - by referencing the values that the LLMs give to items and assessing their relationships. However, GLEs do not attempt to address the nature of the second premise in the Space-LLM conversation - Vector Space – where items are embedded. Vector space is where the manipulation and interpretation of language is enabled. The larger and more aligned the embedding is with the 'words' of the domain of examination, the better the language model is for generating meaning and emerging abilities regarding the domain (Mikolov *et al*. 2013; Han *et al*. 2023). The GLEs assume that, or are trying to determine if, LLMs have a proper vector space for space. GLEs attempt to maximize the existing efficiencies of an insightful, and proven system (Gilson *et al*. 2003). The need for an intrinsic spatial structure is evident when considering that the custom Vector Space built by LLMs is designed based on the data and documents supplied. LLM Vector Spaces are not just containers but shapers of meaning.

The significance of Vector Space is evident when we distinguish between vector database Token Location Embedding (VTLE) and LLM TLE (LTLE). VTLE takes place in the vector space of the language, but LTLE gets meaning from an existing vector space. So, VTLE helps to build and define the language space, whereas the existing language space interprets LTLE. Invoking our Space-LLM conversation, we could say that – the results of VTLE are akin to the native tongue, whereas the results of LTLE are akin to a translation. The native tongue carries intrinsic meaning. Translation tends to carry implied meaning, showing the importance of the vector space of an LLM, per the Space-LLM conversation. Vector Space helps to shape meaning and, hence, what and how the language speaks. The GLEs do not address VTLE. So, GLEs are limited to spatial interpretation. Without being a part of the Vector Space definition, GLEs lack the ability for spatially intrinsic reasoning, which aligns with the proven history of Herring (1991), Goodchild (2009), Kuhn (2012), and Xing and Sieber (2023) that – *spatial insights need to come from a spatial domain.*

Note how the convergence of efficiencies and gaps indicates a *foundational* fix to speaking 'Space.'

**GLE precipitates a Spatial Language Model (SLM) to speak space (Q.3.)**

GLE-LLM improvements and gaps reveal that the Space-LLM conversation only progresses with foundational attention to mode modality and math, as follows:

**LLM:** 1. 'Your vectors must be my vectors' (to enable – mode, modality, and math).
2. 'Your vector space must be my vector space?' (Orients – mode, modality, math).

**Space:** 1. 'My vectors are – {Longitude (Long), Latitude (Lat), and Altitude (Alt)}.'
2. 'My vector space is - Long (-180 to 180), Lat (-90 to 90), Alt (ft).'

**LLM:** 1. 'I will use your vectors to represent items.'
2. 'Your Geometric Coordinate System (GCS) will represent *Vector Space*.'

**Space:** 'Then your type of "knowing" can be my type of "knowing."'
(Here - mode, modality, and math give spatial meaning)

**LLM:** 'Then, I am no longer an LLM; I am a Spatial Foundation/Language Model (SLM).' Note - We use 'Language' in this writing to align with 'Space as a Language (SaaL)' and ‘LLMs,’ hence SLM instead of SFM (Spatial Foundational Model). ‘SFM’ may be suitable for model-specific semantics.

LLMs use vectors in a vector space to establish mode, modalities, and math for language ‘knowing.’ Geo-space uses coordinates in a CGS to establish mode, modalities, and math for spatial ‘knowing.’ A spatial Foundation Model (FM) results when the vector space represents a GCS and the vectors are spatial coordinates. Further, since GLE successes related to the proximity of mode, modality, and math to their domains and analysis, we posit three axioms that supplant the GLE challenges with mode, modality, and math enabling factors:

A.1. *Space Mode* = spatial coordinate system and mode values. (Space and spatial sight).

A.2. Space Modality = composites of simplest modality. (Vector-based representation).

A.3. Space Math = vector math altered for spatial arithmetic. (Gives – spatial reasoning).

With these adjustments, we have a spatially wired LLM brain – SLM, the foundation for a Spatial Artificial Intelligence System (SPAIS). We posit that the model of SLM in Figure 2 models spatial intelligence that allows a computer system to mimic how humans see, represent, and reason about space.

Figure 2. SLM – Vector Space + Space *Mode* (M1) + Space *Modality* (M2) + Spatial *Math* (M3)

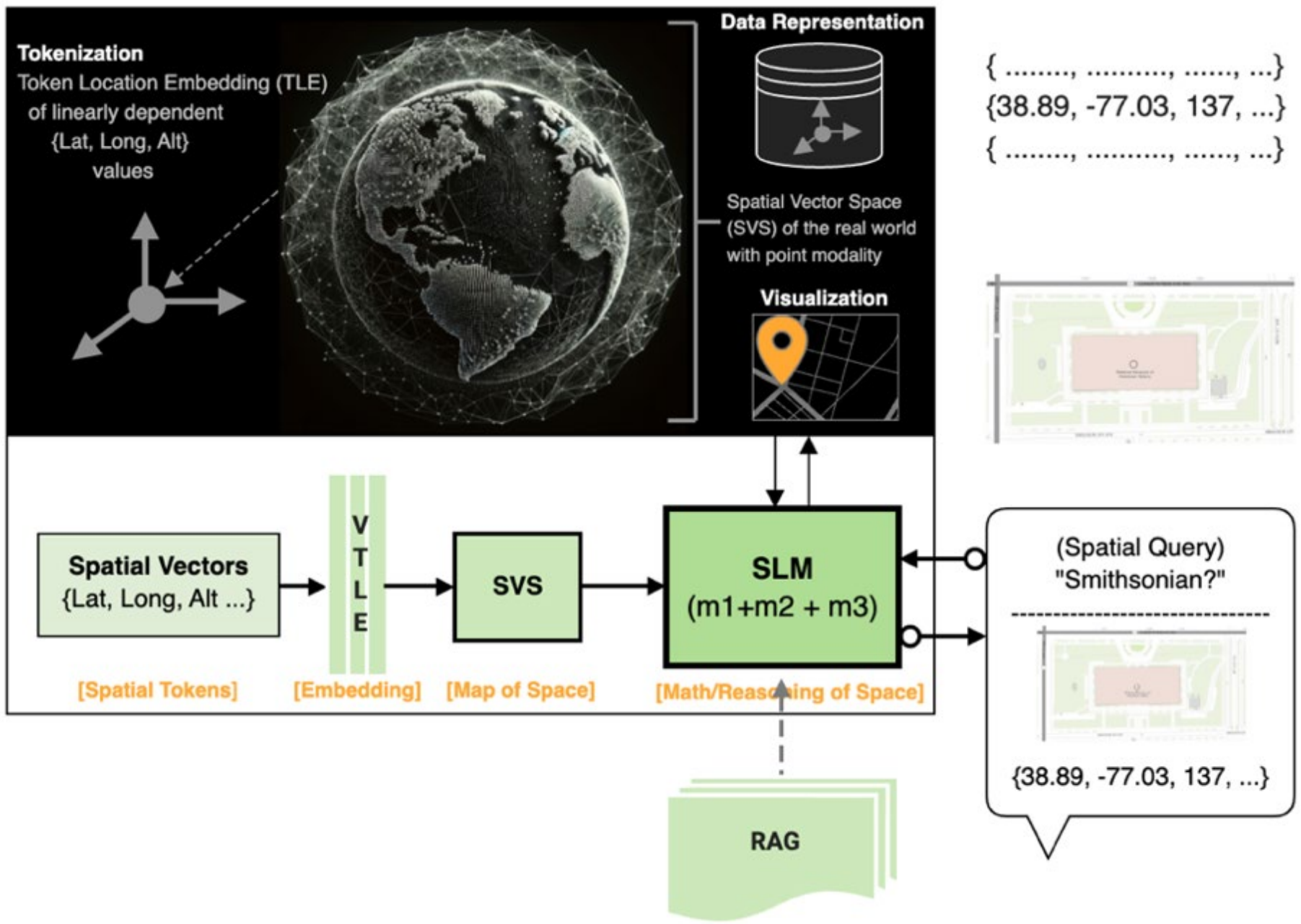

*VTLEs generate a spatially modeled SVS that produces an SLM that uses SVL to reason spatially.*

## SLM illustrated – Spatial mode, spatial modality, and spatial math for deep meaning

Assume Vector Space: Washington, DC.

Assume Location: The Smithsonian National Museum of American History.

A.1.1. First, Mode Values are vectors of spatial coordinates such as Lat, Long, and Alt. For example, an LLM Spatial Vector for the Smithsonian would be {38.890435, -77.030052,137 …}. This coordinate embedding is a Spatial Vector Embedding (SVE).

A.1.2. Space Mode is the frame of the world we are embedding. How do we frame the geospatial? Globally, GCS frames the world as - Long (-180 to 180), Lat (-90 to 90), and Alt (ft).

GCS is an existing Vector Space. We can produce a Vector Space to map GCS using SVEs at a set resolution, producing a grid of linearly dependent coordinates that subdivides our domain and allows us to reference smaller sub-divided independent vectors that represent objects of different modalities (Vijay *et al.* 1986). In doing so, we build a grid that is a model of real space, based on cellular geography (Tobler 1979). This copy of space is a Spatial Vector Space (SVS). SVS is a pre-mapped representation of space at a set resolution where all augmented objects of the same type can use vector mode, modality, and math to map themselves in vector space that maps to real space, using offsets and other factors. The SVS is the 'universe' of space that determines item locations and facilitates modalities and calculations. The ideal is an SVS of as much terrestrial Space as possible. However, since spatial behavior is uniform, representative data size can be relatively small to support unsupervised learning. Such a paradigm aligns with Small Language Models because the required data and model is small, and it extrapolates to the meaning of the entire domain.

Figure 3 shows a representation of SVS extracted from SLM.

Figure 3. Vector space mapping actual space – SVS

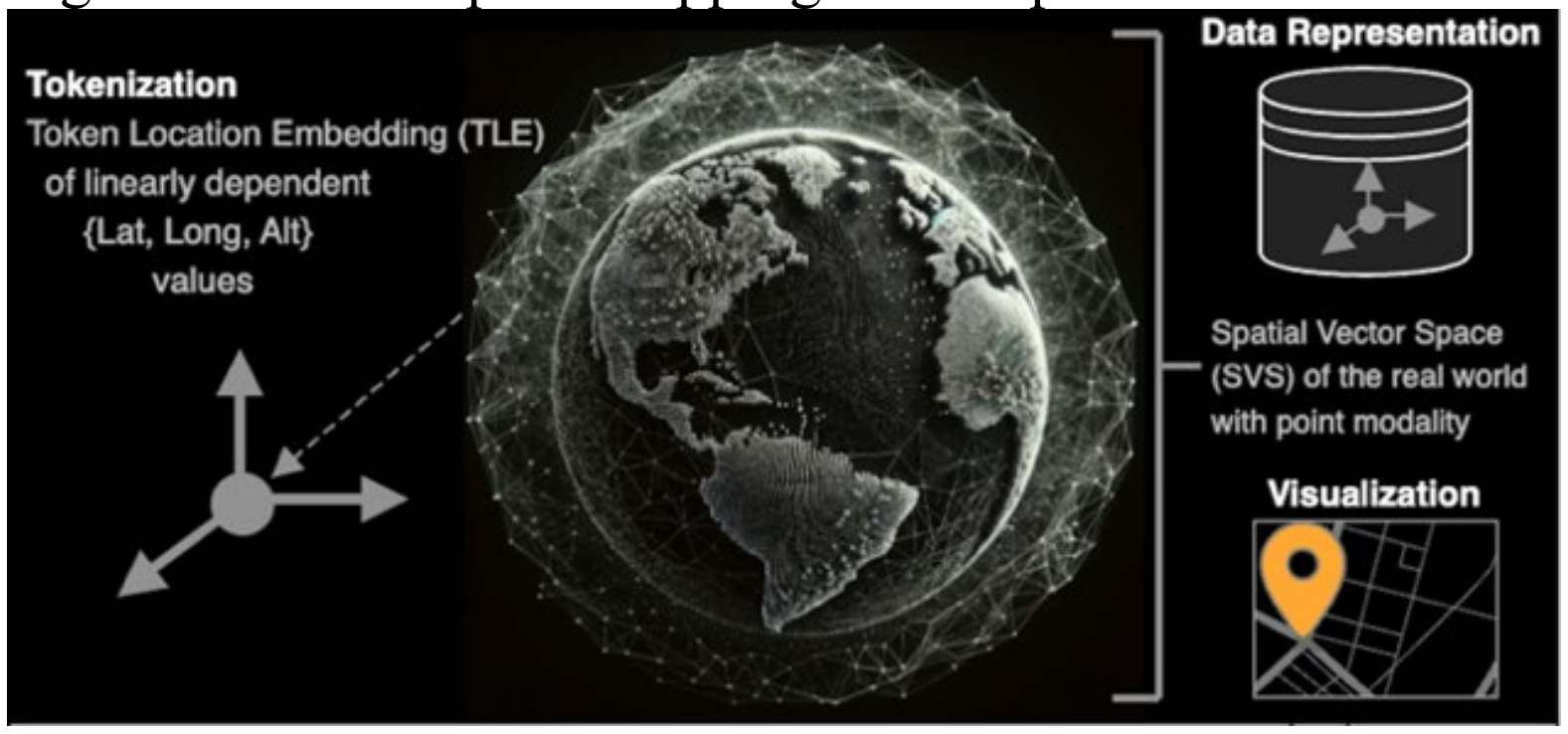

Moving from component vectors to composite *modalities* covering all feature types is challenging.

A.2. Modality is solved by dissolving to and composing from the simplest modality that composes all modalities. This simplest modality is a point. We call Point Modality or Vector Modality (VM). Composing vector modality into more complex modalities is - Composite Vector Modality (CVM). Instinctively, geospatial experts shy away from a massive multipoint database's sheer size and processing needs because it is overwhelming for an RDBMS, but this is manageable for a vector database (Han *et al*. 2023). Table 2 shows the representations, relationships, and modality.

Table 2 – Vector Modality for Spatial Modalities

| **ID** | **Representation** | **Sub-representation** | **Vector Representation** |
|---|---|---|---|
| **1** | ***Points*** will be single vectors | Token Vector 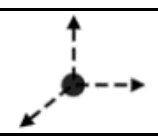 | {38.890435, -77.030052,137…} = Lat, Long, Alt |
| **2** | ***Entities*** are points with location significance. For example – buildings and points of interest | Point 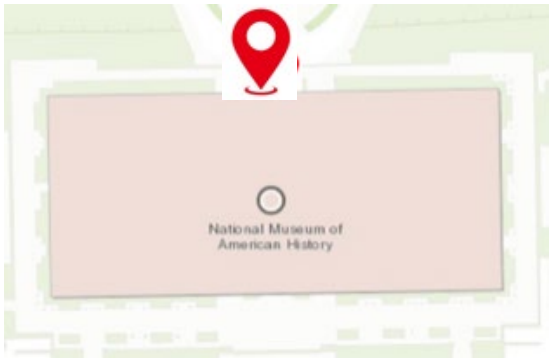 | {38.890435, -77.030052,137, …} |
| **3** | ***Sequences*** will be multiple points that assume lines or arcs between each point. For example – roads and rivers | Points 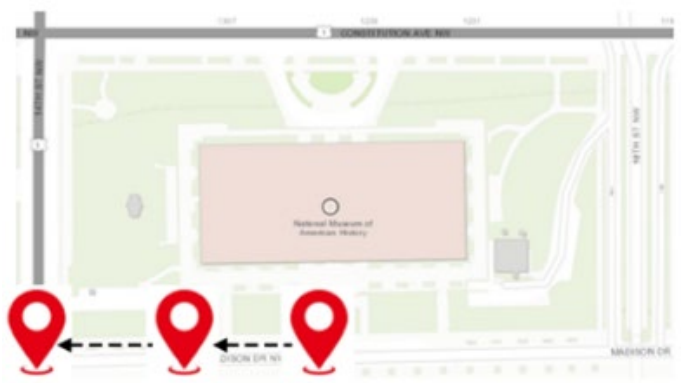 | {38.890435, -77.030052,137, ...3} {38.890426, -77.030931,137, ...3} {38.890408, -77.031959,137, ...3} 3 points along Madison Dr. NW from the Smithsonian to 14th St NW. ID of the sequence (road, path, or another linear feature) is “3” |
| **4** | ***Polygons*** are sequence features that have closure For example – parcel or city boundaries. | Sequences (of points) 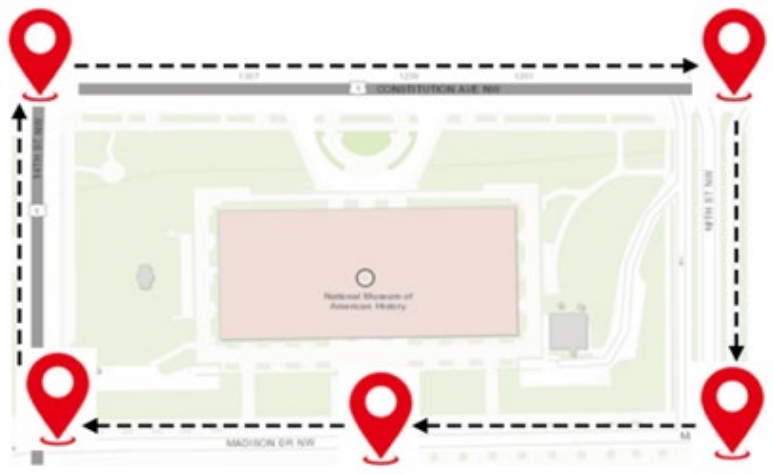 | {38.890435, -77.030052,137, …4} {38.890408, -77.031959,137, …4} {38.892083, -77.031980,137, …4} {38.892086, -77.028097,137, …4} {38.890464, -77.028055,137, …4} {38.890408, -77.031959,137, …4} = a series of points for the road path around the Smithsonian. where polygon ID is “4.” |
| **5** | ***Imagery*** is vector points where each vector also carries the RGB values for its point. For example – satellite imagery or aerial photographs. | Points (of RGBs) 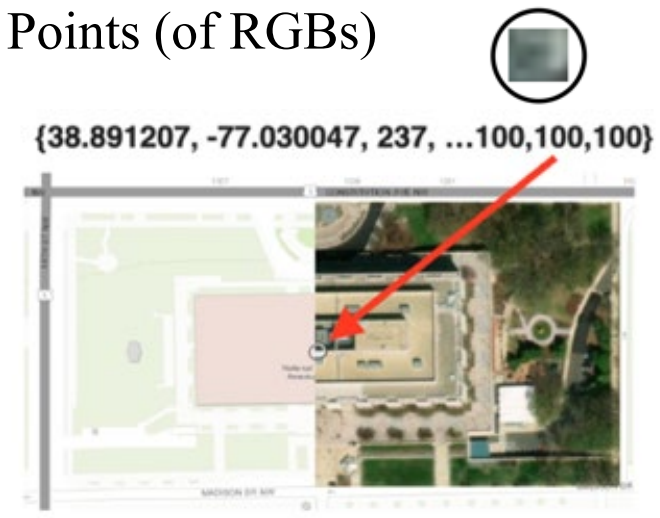  | {38.891207, -77.030047,237, *5*, 100,100,100, …} 100,100,100 is an R, G, B gray section of roof of the Smithsonian. |
| **6** | ***Time series*** features are points sharing a time series ID. For example – traffic or weather data. | Sequences (of closed points) 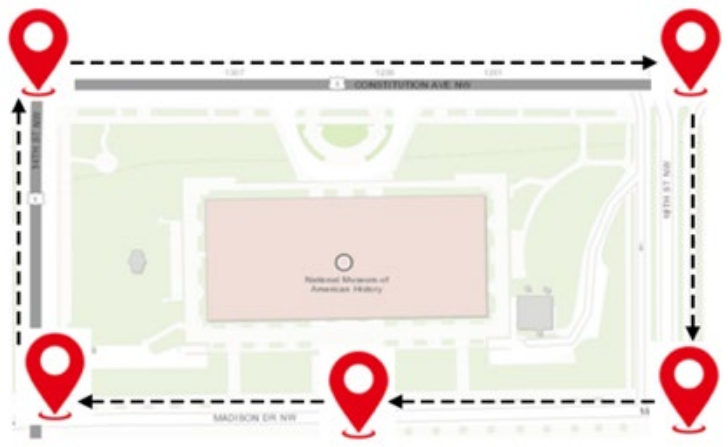 | {38.890435, -77.030052,137 ...*[t1]*} {38.890408, -77.031959,137 ...*[t2]*} {38.892083, -77.031980,137 ...*[t3]*} {38.892086, -77.028097,137 ...*[t4]*} {38.890464, -77.028055,137 ...*[t5]*} {38.890408, -77.031959,137 ...*[t6]*} = a series of points a car arrives at a time *[tx]* around the Smithsonian. |
| **7** | ***Documents*** are points related to entities. For example - maps and news reports related to a feature. | Points (of text) 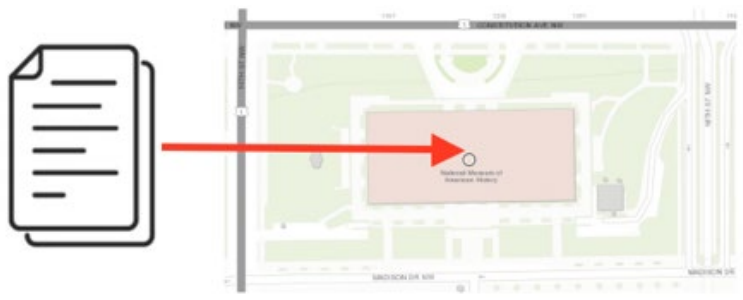 | {38.891207, -77.030047,237 …} Details of Retrieval-Augmented Generation are associated with the point. |

We now require a language that understands the vector space, the vectors, and composite vectors.

A.3. The math of SVS is Spatial Vector Language (SVL). SVL will intuit the vector calculation types for spatial values and relationships. It starts by understanding the state of Space in SVS; for example – were geometric projections already applied in the VTLE? Then, it applies the best vector or augmented calculations for the spatial question and spatial modality. Appendix B has sample recommendations for – distance, spatial overlay, spatial aggregation, nearest neighbor, spatial join, and density analysis in-vector-database-calculations to make SVL work for SVS to represent actual space. Table 2 shows a sample of significant formulas.

Table 2. Sample spatial vector arithmetic

| Spatial Query Type | Best Calculation Type(s) | Explanation |
|---|---|---|
| Distance Query | - Euclidean distance, Geometric (if vectors are 3D) | - Euclidean distance provides simplicity and speed for 2D Space. For 3D data, consider geometric calculations. |
| Nearest Neighbor Query | - Euclidean distance | - Euclidean distance is efficient for finding nearest neighbors in 2D space. |

## Conclusion

The groundbreaking work presented in GLE papers deserves acknowledgment for domain-specific mode, modality, and math improvements in spatial insights. GLE innovations provide insights into the platform and rationale to design beyond the terminal velocity of LLMs. The journey from GLE to the development of Spatial Foundation Models, particularly SLM, marks a paradigm shift in achieving intrinsic spatial awareness. *SLM builds a vector space based on 'actual space' that can represent all modalities and uses vector math for spatial reasoning*. SLM is a foundation model that is space-oriented from the ground up. GLEs inspire the following four-step trajectory for developing spatially intrinsic foundation models: 1. Migration to vector 'spaces' and databases; 2. Adoption of a specialized foundation model mindset; 3. Alignment of SLM and LLMs; and 4. Research steps toward a SPAIS-*ial* approach. *Appendix C* has a listing of related acronyms for GLEs.

### *Migrating to Spatial-Vector Databases (SVD)*

We start with migrating to spatial-vector databases because this is the crawl before the SLM walk that retrains our general mindset to - space as vectors. Ironically, we are rewiring our geospatial technology view to our actual worldview. The move to vectors as an efficient and effective representation of space has been introduced previously. However, it has taken on increased awareness of vector-based imagery by organizations such as the USGS, graph-based geodata, and other hybrid RDBMS + vector data for geospatial uses (Benz *et al*. 2004; Saeedimoghaddam and Stepinski 2020). Vectorized space adoption comes with the incentivization of AI-based insights. Databases such as Weaviate and Milvus are good options to integrate with conventional databases and they are LLM-AI friendly (Lin *et al*. 2023). This also raises the matter of SLM interoperability for seamless integration with current geospatial systems and standards.

*Interoperability – One small step for SLM*

We recommend the following three steps to interoperability with SLM:

(1) *Standardization:* Working closely with standardization bodies to ensure that SLM outputs comply with existing geospatial data formats and standards.
(2) *API Development*: Creating APIs that facilitate the integration of SLMs with existing GIS software and geospatial databases, enabling users to leverage the enhanced spatial reasoning capabilities of SLMs within familiar environments.
(3) *Open-Source Initiatives:* Encouraging the development and sharing of open-source SLM tools and libraries to foster collaboration and adoption across different platforms and domains.

Such databases and interoperability pave a path to *adopting a spatial foundation model mindset*.

**Adopting a spatial foundation model mindset – True representation will set us free**

*Advantages* - SLM will produce intrinsic GeoAI that frees us with native spatial representations that model spatial relationships. GLE furthers the argument of Tomlin (1990) that 'vector seems more *correcter*' than raster. Thus, we can achieve – real-time twin cities and urban planning models, live spatiotemporal feedback, authentic augmented reality, and real-world geocoding. It will also set a precedent for other space-vector fields, to implement Intrinsic LMs (ILMs) that map foundational domain representations for deeper 'knowing.' Integrating spatial vector databases with existing systems will provide superior spatial representation in mode, modality, and reasoning (Brinkhoff 2003; Dunfey *et al*. 2006). Correct representations of space give us accurate insights—APIs, frameworks, and initiatives that build a correct vector-based mode-modality-math foundation. SLM correctness needs to be measurable in terms of ability and scalability.

*SLM ability* should be evaluated in terms of four factors - accuracy, complexity, efficiency, and scalability as follows:

(1) *Geospatial Accuracy:* Determine how precisely SLMs can identify and describe geographical locations, boundaries, and features compared to ground truth data.
(2) *Reasoning Complexity:* How well can SLMs handle complex spatial queries and provide contextually relevant responses?
(3) *Efficiency Metrics:* Benchmark relative speed of query processing, memory usage, and energy consumption, particularly when handling large geospatial datasets.
(4) *Scalability Tests:* Performance evaluations on incrementally increased dataset sizes to assess the model's capacity to maintain efficiency and accuracy at scale. The complexity of geospatial data and queries means special consideration must be given to SLM scalability.

*SLM scalability* in AI requires –distributed processing and scalable learning.

(1) *Distributed Processing:* Leveraging distributed computing architectures to parallelize vector space calculations, enabling the SLMs to handle large-scale geospatial data and allow LLM integration.

(2) *Incremental Learning:* Exploring methods for SLMs to learn incrementally from new geospatial data in a RAG type format, ensuring the model remains current and scalable beyond the SVS.

## Integrating SLM and LLMs – Coordination via a Large Language Nexus (LLN)

GLE shows that we are ready to move beyond force-feeding 'space' to LLMs. Trying to force all meaning into an LLM vector space and logic is like trying to have a single-hemisphere brain. What if our brains had no specialization? What if it were a single structure trying to be perfect at everything, about everything, to answer everything? We would not be very insightful nor efficient. On the other hand, Kanwisher (2010) connects distributed cognition to intelligence (Ralph *et al*. 2016). Das (2023) deduced - we need to design humanly - from a human-centric perspective. So, for AI, we need distributed domain specific mode, modality, and math to augment natural language intelligence with spatial and other specialized intelligence.

We posit RAG-based DLEs as a foundation for bridging an SLM with LLMs via a Large Language Nexus (LLN). This approach parallels the brain's corpus callosum and envisions a synergy between the spatially aware SLM and reasoning-centric LLMs. SLM will be like the right hemisphere - spatially aware, and LLMs like the left hemisphere - reasoning, articulation, and comprehension. Figure 4 illustrates the SLM-LLN-LLM.

Figure 4. SLM and LLM integrated via a Large Language Nexus (LLN)

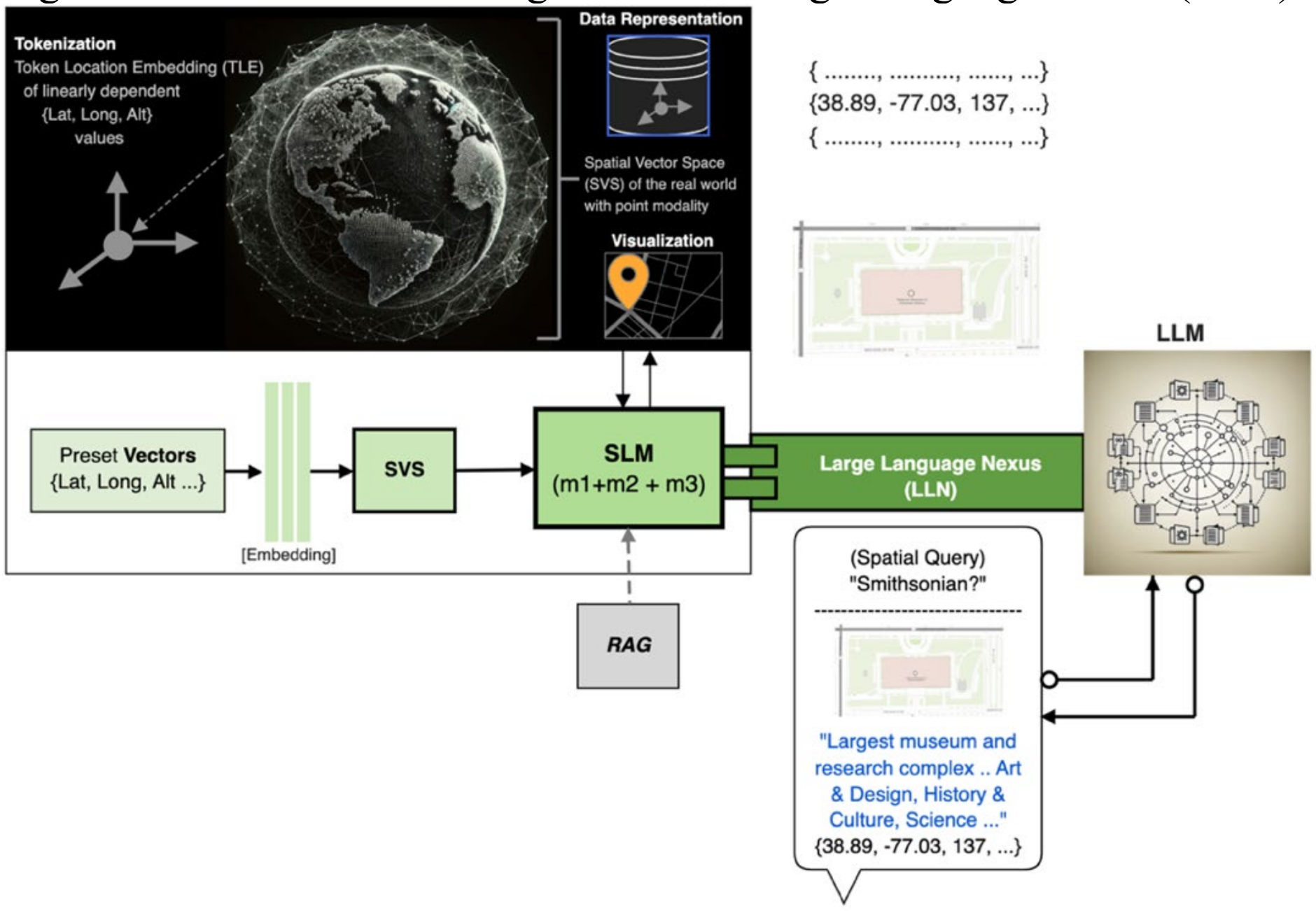

A hypothetical Large Language Nexus (LLN) manages Distributed Foundation Modeling (DFM) by channeling assignments and unified knowledge for cohesive representation (compare Figure 3 and Figure 4), hierarchy, context, and other executive factors to synergistically model rich spatial reasoning and expressions.

To truly emulate human intelligence, the call to design distributed language models, accommodating domain-specific modes, modalities, and math resonates with GLE research and innovation. GLEs prove domain expectations should match domain vector space and reasoning, as per SVS.

SVS sets a precedent for Foundation Vector Spaces (FVS) for deep domain representation. SLMs will set a precedent for Specialized Foundation Models based on FVSs that facilitate distributed LMs imbued with domain-specific insights. Such specialization will facilitate distributed data and training loads for domain generalization and unsupervised learning. SLM will allow us to reason spatially, and SLM-LLN-LLM will allow us to connect and converse 'Spatially.'

SLM/SLM-LLN-LLM require a clear trajectory via clear research steps to arrive at SPAIS.

## Research steps to SPAIS – GeoAI needs its own space

GeoAI needs to carve out a space for *space* through research of SLM mode, modality, and math via 1. Spatial VTLE and SVS, 2. Vectorized Modality, and 3. SVL. *VTLE forms an SVS that uses SVL to speak space as a language*, giving us a Spatial AI System (SPAIS) or true GeoAI. VTLE is pivotal, the cornerstone, construct, and the components of the Spatial brain. VTLE will facilitate the downstream representation of composite vector features and 'pointed' analysis. This GLE reveals the great insight of the Geospatial-LLM corpus that points us to specialized spatial foundation models for spatial 'knowing (Ling *et al*. 2023). GLEs are also nudging us to be reinspired by human intelligence to design distributed language models in the form of foundation modules (Kanwisher 2010).

A sample of the research trajectory for spatial foundation modeling and integration with LLMs is as follows:

- 'Token Location Embedding for Spatially Aware Databases and Vector Spaces.'
- 'From Vector Modality to Composite Vector Modality: Representing Spatial Modalities.'
- 'Spatial Math for Vector Space: A Path to Spatial AI Systems (SPAIS).'
- 'Large Language Nexus: Dissecting and Connecting the Hemispheres of AI Models.'

… paving the way for *SPAIS.*

## Data and codes availability statement

https://figshare.com/s/2ba00dd431355bd6bc1c

## References

Balsebre, P., Huang, W., Cong, G., Li, Y., 2023. CityFM: City foundation models to solve urban challenges. arXiv preprint arXiv:2310.00583.

Batty, M., Xie, Y., 1994. From cells to cities. *Environment and Planning B,* 21 (7), S31–S48.

Batty, M., Xie, Y., Sun, Z., 1999. Modeling urban dynamics through GIS-based cellular automata. Computers, *Environment and Urban Systems*, 23 (3), 205–233.

Benz, U.C., Hofmann, P., Willhauck, G., Lingenfelder, I., Heynen, M., 2004. Multi-resolution, object-oriented fuzzy analysis of remote sensing data for GIS-ready information. *ISPRS Journal of Photogrammetry and Remote Sensing*, 58 (3–4), 239–258.

Bhandari, P., Anastasopoulos, A., Pfoser, D. 2023. Are large language models geospatially knowledgeable? arXiv. Cornell University.

Brinkhoff, T., 2003. A portable SVG-viewer on mobile devices for supporting geographic applications. http://iapg.jade-hs.de/personen/brinkhoff/paper/AGILE2003.pdf. In Proceedings of the 6th Agile Conference on Geographic Information Science, Lyon, France (pp. 87–96).

Das, S., 2023. *Evaluating the capabilities of large language models for spatial and situational understanding* Thesis (MA). University of Cambridge.

Dunfey, R.I., Gittings, B.M., Batcheller, J.K., 2006. Towards an open architecture for vector GIS. *Computers and Geosciences*, 32 (10), 1720–1732.

Fernandez, A., Dube, S., 2023. Core building blocks: Next gen geo spatial GPT application. arXiv. Cornell University.

Goodchild, M.F., 2009. Geographic information systems and science: Today and tomorrow. *Annals of GIS*, 15 (1), 3–9.

Goodchild, M.F., Janelle, D.G., 2004. Thinking spatially in the social sciences. *Spatially Integrated Social Science*, 3–17.

Han, Y., Liu, C., Wang, P., 2023. A comprehensive survey on vector database: Storage and retrieval technique, challenge. arXiv. Cornell University.

Herring, J.R., 1991. The mathematical modeling of spatial and nonspatial information in geographic information systems. In Cognitive and Linguistic Aspects of Geographic Space, (313–350).

Hu, Y., Mai, G., Cundy, C., Choi, K., Lao, N., Liu, W., Lakhanpal, G., Zhou, R.Z., Joseph, K., 2023. Geo-knowledge-guided GPT models improve the extraction of location descriptions from disaster-related social media messages. *International Journal of Geographical Information Science*, 37 (11), 2289–2318.

Janowicz, K., Gao, S., McKenzie, G., Hu, Y., Bhaduri, B., 2020. GeoAI: Spatially explicit artificial intelligence techniques for geographic knowledge discovery and beyond. *International Journal of Geographical Information Science*, 34 (4), 625–636.

Ji, Y., Gao, S., 2023. Evaluating the effectiveness of large language models in representing textual descriptions of geometry and spatial relations. arXiv. Cornell University.

Jin, M., Wen, Q., Liang, Y., Zhang, C., Xue, S., Wang, H., Zhang, J., Hu, X., Chen, H., Li, X., Pan, S., Tseng, V.S., Zheng, Y., Chen, L., Xiong, H., 2023. Large models for time series and spatiotemporal data: A survey and outlook. arXiv (Cornell University).

Kanwisher, N., 2010. Functional specificity in the human brain: A window into the functional architecture of the mind. *Proceedings of the National Academy of Sciences of the United States of America*, 107 (25), 11163–11170.

Kuhn, W., 2012. Core concepts of spatial information for transdisciplinary research. *International Journal of Geographical Information Science*, 26 (12), 2267–2276.

Leszczynski, A., Crampton, J.W., 2016. Introduction: Spatial big data and everyday life. *Big Data and Society*, 3 (2).

Li, Z., Zhou, W., Chiang, Y., Chen, M., 2023. GeoLM: Empowering Language Models for Geospatially Grounded Language Understanding. arXiv (Cornell University).

Lin, J., Pradeep, R., Teofili, T., Xian, J., 2023. Vector search with OpenAI embeddings: Lucene is all you need. arXiv. Cornell University.

Ling, C., Zhao, X., Lu, J., Deng, C., Zheng, C., Wang, J., Chowdhury, T., Li, Y., Cui, H., Zhang, X., Zhao, T., Panalkar, A., Cheng, W.S., Wang, H., Liu, Y., Chen, Z., Chen, H., White, C.M., Gu, Q., . . . Zhao, L., 2023. Domain specialization as the key to make large language models disruptive: A comprehensive survey. arXiv. Cornell University.

Liu, P., Biljecki, F., 2022. A review of spatially explicit GeoAI applications in urban geography. *International Journal of Applied Earth Observation and Geoinformation*, 112, 102936.

Luo, S., Ni, J., Chen, S., Rong, Y., Xie, Y., Liu, L., Jin, Z., Yao, H., Jia, X., 2023. The foundational semantic recognition for modeling environmental ecosystems. arXiv. Cornell University.

Mai, G., Cundy, C., Choi, K., Hu, Y., Lao, N., Ermon, S., 2022. Towards a foundation model for geospatial artificial intelligence (vision paper). In Proceedings of the 30th International Conference on Advances in Geographic Information Systems, 1–4.

Mai, G., Huang, W., Sun, J., Song, S., Mishra, D., Liu, N., Gao, S., Liu, T., Cong, G., Hu, Y., Cundy, C., Li, Z., Zhu, R., Lao, N., 2023. On the opportunities and challenges of foundation models for geospatial artificial intelligence. arXiv. Cornell University.

Manvi, R.S., Khanna, S., Mai, G., Burke, M., Lobell, D.B., Ermon, S., 2023. GeoLLM: Extracting geospatial knowledge from large language models. arXiv. Cornell University.

Mikolov, T., Chen, K., Corrado, G., Dean, J., 2013. Efficient estimation of word representations in vector space. arXiv. https://arxiv.org/pdf/1301.3781. Cornell University.

Openshaw, S., Openshaw, C., 1997. Artificial intelligence in geography. John Wiley & Sons, Inc.

Saeedimoghaddam, M., Stepinski, T.F., 2020. Automatic extraction of road intersection points from USGS historical map series using deep convolutional neural networks. *International Journal of Geographical Information Science*, 34 (5), 947–968.

Salmas, K., Pantazi, D.A., Koubarakis, M., 2023. Extracting geographic knowledge from large language models: An experiment.

Smith, T.R., 1984. Artificial intelligence and its applicability to geographical problem solving. *Professional Geographer*, 36 (2), 147–158.

Tobler, W.R., 1979. Cellular geography. *Philosophy and Geography*, 379–386.

Tomlin, C.D., 1990. Geographic information systems and cartographic modeling, 249. Prentice Hall.

Unlu, E., 2023. Chatmap: Large Language Model interaction with cartographic data. arXiv. Cornell University.

Xing, J., Sieber, R., 2023. The challenges of integrating explainable artificial intelligence into GeoAI. *Transactions in GIS*, 27 (3), 626–645.

Yan, Y., Wen, H., Zhong, S., Chen, W., Chen, H., Wen, Q., Zimmermann, R., Liang, Y., 2023. When urban region profiling meets large language models. arXiv. Cornell University.

Yin, Z., Li, D., Goldberg, D.W., 2023. Is ChatGPT a game changer for geocoding – a benchmark for geocoding address parsing techniques. In Proceedings of the 2nd ACM SIGSPATIAL International Workshop on Searching and Mining Large Collections of Geospatial Data (pp. 1–8).

**Appendix A. GLE – Theme selection and final papers**

Table A.1 – GLE Theme Selection Criteria

| ***GLEs*** | **ELE** | **DLE** | **SLE** | **TLE** |
|---|---|---|---|---|
| **Theme cognates** | Entity, location, modality, toponym | Documents, RAG, OSM | Modality, network, sequence, timeseries | Tokenize, geometry |

Table A.2 GLE – Final papers and associated embeddings

| **Paper** | **ELE** | **DLE** | **SLE** | **TLE** | **SVS** |
|---|---|---|---|---|---|
| Towards a foundation model for geospatial artificial intelligence (vision paper) (Mai et al., 2022) | x | | x | | |
| On the opportunities and challenges of foundation models for geospatial artificial intelligence (Mai et al., 2023) | x | | x | | |
| GeoLLM: Extracting geospatial knowledge from large language models (Manvi et al., 2023) | x | x | x | | |
| Chatmap: Large language model interaction with cartographic data (Unlu., 2023) | x | x | x | | |
| Are large language models geospatially knowledgeable? (Bhandari et al., 2023) | x | x | x | | |
| CityFM: City foundation models to solve urban challenges (Balsebre et al., 2023) | x | x | x | | |
| Core building blocks: Next gen geospatial GPT application (Fernandez and Dube, 2023) | x | | x | x | |
| GeoLM: Empowering language models for geospatially grounded language understanding (Li et al., 2023) | x | x | x | x | |
| Extracting geographic knowledge from large language models: An experiment (Salmas et al., 2023) | x | | x | | |
| Evaluating the effectiveness of large language models in representing textual descriptions of geometry and spatial relations (Ji and Gao, 2023) | x | x | x | x | |
| When urban region profiling meets large language models (Yan et al., 2023) | x | | x | x | |
| Domain specialization as the key to make large language models disruptive: A comprehensive survey (Ling et al., 2023) | x | | x | x | |
| Large models for time series and spatio-temporal data: A survey and outlook (Jin et al., 2023) | x | x | x | | |
| FREE: The foundational semantic recognition for modeling environmental ecosystems (Luo et al., 2023) | x | x | x | | |
| Evaluating the capabilities of Large Language Models for spatial and situational understanding (Das, 2023) | x | | x | | |
| Is ChatGPT a game changer for geocoding--a benchmark for geocoding address parsing techniques (Yin et al., 2023) | x | x | x | | |

**Appendix B - Spatial vector language (SVL) Table**

This SVL Table – B.1, provides general guidance on the principles of calculation for various spatial query types. It includes explanations for each choice based on factors like simplicity, speed, accuracy, and the specific nature of the query. This illustrates the principle of SVL. It is not a comprehensive list.

Table B.1. – SVL Queries, Calculation Types, and Explanations

| Spatial Query Type | Best Calculation Type(s) | Explanation |
|---|---|---|
| Distance Query | - Euclidean distance, Geometric (if vectors are 3D) | - Euclidean distance provides simplicity and speed for 2D space. For 3D data, consider geometric calculations. |
| | - Haversine formula (great-circle distance) | - Haversine formula is suitable for geographic coordinates, accounting for Earth's curvature. |
| | - Vincenty's formulae | - Vincenty's formulae offer high precision for geographic distances over the Earth's surface. |
| | - Coordinate-based representation | - Utilizes spatial coordinates for precise distance measurements. |
| Spatial Overlay Query | - Geometric calculations | - Geometric calculations are suitable for determining spatial relationships between vector geometries. |
| | - Euclidean calculations (if vectors are 3D) | - Euclidean calculations can be considered for vector representations modeling 3D space. |
| Spatial Aggregation Query | - Arithmetic calculations | - Arithmetic calculations, such as summation and averaging, are suitable for aggregating attribute values within geometries. |
| | - Geometric calculations (if required for spatial aggregation) | - Geometric calculations may be used when spatial attributes are involved in aggregation operations. |
| | - Custom aggregation methods (if needed) | - Custom aggregation methods allow specific rules for data aggregation within spatial features. |
| Nearest Neighbor Query | - Euclidean distance | - Euclidean distance is efficient for finding nearest neighbors in 2D space. |
| | - Haversine formula (great-circle distance) | - Haversine formula is suitable for geographic coordinates, accounting for Earth's curvature. |
| Spatial Join Query | - Geometric calculations | - Geometric calculations are adequate for determining spatial relationships and overlaps between vector features. |
| | - Euclidean calculations (if vectors are 3D) | - Euclidean calculations used for 3D spatial join operations for 3D space. |
| Density Analysis Query | - Arithmetic calculations | - Arithmetic calculations can be applied for density analysis by aggregating data within specified areas. |
| | - Geometric calculations (for density analysis) | - Geometric calculations may be used when spatial attributes influence density analysis. |

**Appendix C. SPAIS acronyms and relationships**

| Acronym | Full Term | Related Acronyms | Explanation/Relationships |
|---|---|---|---|
| CVM | Composite Vector Modality | VM, SLM, SVL | CVM is a logical grouping of VMs that identify a spatial unit. |
| DLE | Document Location Embedding | GLE, ELE, SLE, TLE | DLE, ELE, SLE, and TLE are types of GLE that focus on embedding different types of geospatial data into LLMs. |
| ELE | Entity Location Embedding | GLE, DLE, SLE, TLE | DLE, ELE, SLE, and TLE are types of GLE that focus on embedding different types of geospatial data into LLMs. |
| GLE | Geospatial Location Embedding | DLE, ELE, SLE, TLE, SLM, SVL | GLE is the overarching term for embedding geospatial data into LLMs. |
| LLN | Large Language Nexus | LLM, SLM | LLN is a Nexus that allows synergistic Language Model collaboration. |
| SLM | Spatially Language Model | TLE | SLM is an AI model produced by a TLE that embeds geospatial coordinates as a map of space and is used to create a vector space that uses spatial modes, modality, and math spatial reasoning. |
| SPAIS | Spatial AI System | SLM, SVL, LLM | SPAIS is an SLM using SVL for spatial reasoning. The SLM may be integrated with an LLM via an LLN. |
| SVL | Spatial Vector Language | VTLE, SVS, SLM | SVL is the math and logic corpus used to *reason* on an SLM using its SVS. |
| SVS | Spatial Vector Space | VTLE, SLM, SVL | SVS is a representation of real space using TLEs mapped at a set resolution. |
| TLE | Token Location Embedding | GLE, DLE, ELE, SLE | TLE is a type of GLE that embeds geospatial data into LLMs by assigning locations to individual tokens. |
| VM | Vector Modality | HSWN, VTLE, SVS | VM is a modality represented by a single vector or point. |
| VTLE | Vector Token Location Embedding | SVS, SLM, SVL | VTLE is tokenization of spatial values as dimensions in a Language Model |